\documentclass[twocolumn,floatfix,preprintnumbers,nofootinbib,superscriptaddress]{revtex4}

\usepackage{ulem}
\usepackage{bm}
\usepackage{times}
\usepackage{amssymb,amsbsy,amsmath,amsfonts}
\usepackage{graphicx}
\usepackage{float}
\usepackage{color}
\usepackage{morefloats}
\usepackage{rotating}
\usepackage{srcltx}
\usepackage{slashed}
\usepackage{subfigure}
\usepackage{multirow}
\usepackage{verbatim}
\usepackage{hyperref}
\usepackage{tabularx}
\usepackage{adjustbox}
\usepackage{array}

\begin{document}
	
\title{The dynamically generated $N(1535)$ state in the $\Lambda_c^+ \to p\bar{K}^0 \pi^0$ decay} 
	
	\author{Ying Li}
    \affiliation{School of Physics, Beihang University, Beijing 102206, China}
    \affiliation{School of Physics, Zhengzhou University, Zhengzhou 450001, China}
	
	\author{En Wang}~\email{wangen@zzu.edu.cn}
	\affiliation{School of Physics, Zhengzhou University, Zhengzhou 450001, China}
	
\author{Li-Sheng Geng}~\email{lisheng.geng@buaa.edu.cn}
 \affiliation{Sino-French Carbon Neutrality Research Center, \'{E}cole Centrale de P\'{e}kin/School
of General Engineering, Beihang University, Beijing 100191, China}
\affiliation{School of Physics, Beihang University, Beijing 102206, China}
\affiliation{Peng Huanwu Collaborative Center for Research and Education, Beihang University, Beijing 100191, China}
\affiliation{Beijing Key Laboratory of Advanced Nuclear Materials and Physics, Beihang University, Beijing 102206, China }
\affiliation{Southern Center for Nuclear-Science Theory (SCNT), Institute of Modern Physics, Chinese Academy of Sciences, Huizhou 516000, China}
\vspace{0.5cm}

\author{Ju-Jun Xie}~\email{xiejujun@impcas.ac.cn}
\affiliation{Southern Center for Nuclear-Science Theory (SCNT), Institute of Modern Physics, Chinese Academy of Sciences, Huizhou 516000, China}
\affiliation{State Key Laboratory of Heavy Ion Science and Technology, Institute of Modern Physics, Chinese Academy of Sciences, Lanzhou 730000, China} 
\affiliation{School of Nuclear Sciences and Technology, University of Chinese Academy of Sciences, Beijing 101408, China}

\vspace{0.5cm}
\date{\today}
	
	\begin{abstract}

We present a theoretical analysis of the process $\Lambda_c^+ \to p\bar{K}^0 \pi^0$ within the chiral unitary approach, with particular emphasis on the dynamically generated $N(1535)$ resonance. In addition to $N(1535)$, our model incorporates contributions from other intermediate resonances including $N(1650)$, $K^*(892)$, $K_0^*(1430)$, $N(1440)$, and $\Sigma(1750)$. The calculated invariant mass distributions and Dalitz plot are in good agreement with the recent Belle measurements. Our analysis highlights the crucial role of $N(1535)$ state in this decay channel and supports its interpretation as a dynamically generated state arising from coupled-channel meson-baryon interactions.

	\end{abstract}
	
	\maketitle
	
	\section{Introduction} \label{sec:Introduction}

Multibody hadronic decays of charmed baryons offer valuable insights into the properties of light hadron resonances, given their large phase space and significant final-state interactions~\cite{Wang:2024jyk,Li:2025exm,Feng:2020jvp,Zeng:2020och,Wang:2020pem,Xie:2017erh,Xie:2017xwx}. For instance, the $\Lambda(1670)$ resonance, with quantum numbers $J^P = 1/2^-$, manifests as a dip structure near the $\eta\Lambda$ threshold in the $\bar{K}^0n$ invariant mass distribution of the $K^- p \to \bar{K}^0 n$ scattering~\cite{Gopal:1976gs}, and as an enhancement in the $\eta\Lambda$ invariant mass distribution in both $K^-p\to \eta\Lambda$~\cite{CrystalBall:2001uhc} and $\Lambda_c^+\to \eta\Lambda\pi^+$~\cite{BESIII:2018qyg,Belle:2020xku,BESIII:2024mbf,Lyu:2024qgc,Duan:2024czu}, suggesting that $\Lambda(1670)$ may possess an exotic nature.

In 2023, the LHCb Collaboration performed an amplitude analysis of $\Lambda^+_c \to p K^- \pi^+$  in connection with semileptonic beauty hadron decays~\cite{LHCb:2022sck}, observing a cusp structure near the $\eta\Lambda$ threshold in the $K^- p$ invariant mass distribution. Subsequently, the Belle Collaboration also measured $\Lambda^+_c \to pK^-\pi^+$ and found a narrow peak near the $\eta\Lambda$ threshold~\cite{Belle:2022cbs}. Recent theoretical studies of $\Lambda^+_c \to pK^-\pi^+$ within the chiral unitary approach support the interpretation of this cusp as arising from the $\Lambda(1670)$ pole~\cite{Zhang:2024jby,Duan:2024okk}.

Recently, the Belle Collaboration reported a precise measurement of the branching fraction ratio $\mathcal{B}(\Lambda_c^+\to pK_S^0\pi^0)/\mathcal{B}(\Lambda_c^+\to pK^-\pi^+) = 0.339 \pm 0.002\pm0.009$ using $980~\text{fb}^{-1}$ of $e^+e^-$ collision data, and presented the first investigation of intermediate resonances in the decay $\Lambda_c^+\to pK_S^0\pi^0$~\cite{Belle:2025voy}. In Fig.~7 (c) of that work, two distinct peak structures are visible in the $M_{p \pi^0}$ invariant mass distribution, which could be associated with the $N(1535)$ and $N(1650)$ resonances. However, Belle did not perform a full amplitude analysis, which would be essential to clarify the contributions of these intermediate resonances.

The peak structure near the $p\eta$ threshold corresponds to the diagonal band observed in the Dalitz plot (left panel of Fig.~6 in Ref.~\cite{Belle:2025voy}), likely attributable to a threshold cusp enhanced by the $N(1535)$ resonance. A similar effect was observed in the study of $\Lambda_c^+\to pK_S^0\eta$~\cite{Belle:2022pwd,Li:2024rqb}, which resembles the $\Lambda\eta$ threshold cusp enhanced by $\Lambda(1670)$ in the $M_{pK^-}$ distribution~\cite{Zhang:2024jby,Duan:2024okk}. Additionally, a peak structure appears around $1650$~{MeV} in the $M_{p\pi^0}$ distribution~\cite{Belle:2025voy}, associated with the $N(1650)$ contribution. 

As required by isospin symmetry, the production of the $\Delta^{++}K^-$ channel is favored over the $\Delta^+\bar{K}^0$ channel~\cite{Hsiao:2020iwc,Savage:1989qr}. Thus, compared to $\Lambda_c^+\to pK^-\pi^+$, the $\Delta$ contribution in $\Lambda_c^+\to p K_S^0\pi^0$ is suppressed. This is consistent with the relative strengths of $\Delta(1232)$ shown in Figs.~7(c) and (d) of Ref.~\cite{Belle:2025voy}. The process $\Lambda_c^+\to p K_S^0\pi^0$ is therefore more favorable for studying nucleon excited states.

The internal structure of $N(1535)$ ($J^P=1/2^-$) remains an open question. Within the constituent quark model, it presents two puzzles: first, the mass inversion problem relative to the $J^P = 1/2^+$ radial excitation $N(1440)$ and the strangeness $S = -1$ state $\Lambda(1405)$~\cite{Capstick:2000qj}; second, its unexpectedly large coupling to strangeness channels such as $\eta N$, $\eta' N$, $K\Lambda$, and $K\Sigma$~\cite{Liu:2005pm,Geng:2008cv,CLAS:2005rxe,Cao:2008st}, which suggests a significant $s\bar{s}$ component in the $N(1535)$ wave function.

Refs.~\cite{Helminen:2000jb,Zou:2007mk,Hannelius:2000gu} have proposed interpreting $N(1535)$ as a three-quark excited state with orbital quantum number $L = 1$, admixed with a $[ud][us]\bar{s}$ pentaquark component. This hybrid picture can naturally explain both the mass of $N(1535)$ and its strong coupling to strangeness channels. Numerous studies have also investigated the molecular nature of $N(1535)$. In Refs.~\cite{Kaiser:1995eg,Kaiser:1996js}, $N(1535)$ was interpreted as a bound state of $K\Lambda$ and $K\Sigma$ using the chiral unitary approach. Subsequent studies~\cite{Nieves:2001wt,Bruns:2019fwi,Nacher:1999vg,Kaiser:1995eg,Bruns:2010sv,Khemchandani:2013nma,Inoue:2001ip,Nieves:2011gb,Gamermann:2011mq,Kaiser:1996js,Garzon:2014ida,Li:2024rqb,Li:2025gvo,Lyu:2023aqn} have further explored and supported this molecular picture description within coupled-channel frameworks.

Recently, studies have tested the molecular nature of $N(1535)$ through measurements of its correlation functions~\cite{Molina:2023jov,Liu:2025eqw} or scattering lengths and effective ranges of $K\Sigma$, $K\Lambda$, and $\eta p$ channels~\cite{Li:2023pjx}. Within Hamiltonian effective field theory, $N(1535)$ has been interpreted as primarily a three-quark state dressed by $\pi N$ and $\eta N$ interactions~\cite{Liu:2015ktc,Guo:2022hud,Abell:2023nex}.


In this work, we adopt the chiral unitary approach to investigate the properties of $N(1535)$ in the process $\Lambda_c^+\to p \bar{K}^0\pi^0$. We dynamically generate $N(1535)$ through $S$-wave pseudoscalar meson-octet baryon interactions, while also considering contributions from $K^*(892)$, $K_0^*(1430)$, $N(1650)$, and other possible intermediate resonant states. Our results show good agreement with the Belle invariant-mass distributions and Dalitz plot.

	\section{Formalism} \label{sec:Formalism}
	

		\begin{figure}[tbhp]
		\centering
		\includegraphics[scale=0.55]{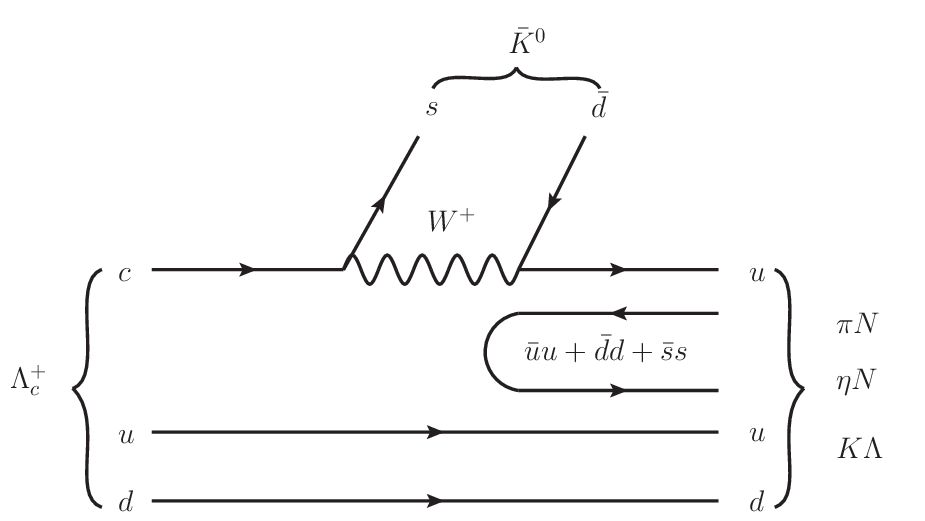}
		\caption{Quark level diagram for the $\Lambda_c^+\to\bar{K}^0PB$ with hadronization of the $uud$ pair.}
		\label{fig:quark}
	\end{figure}

\begin{figure}[tbhp]
		\centering
		\includegraphics[scale=0.8]{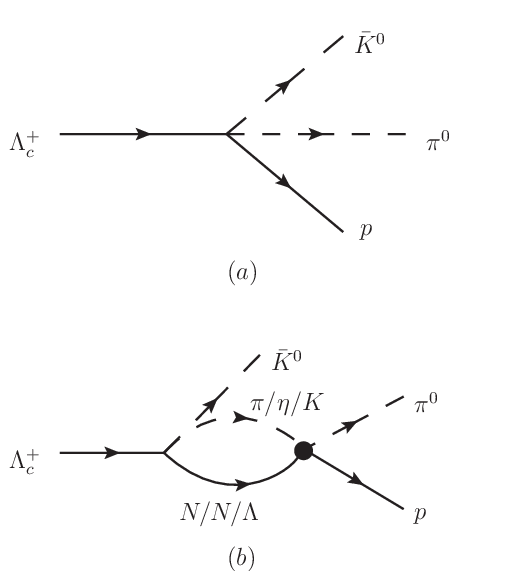}
		\caption{Hadron-level diagrams for $\Lambda_c^+ \to p\bar{K}^0 \pi^0$ decay, including the tree level (a) and final state interactions (b). The blob represents the meson-baryon scattering amplitude $t_{MB\to\pi^0 p}$.}
		\label{fig:hadron1690}
	\end{figure}

 We begin by discussing the primary vertex of the $\Lambda_c^+ \to p\bar{K}^0 \pi^0$ decay at the quark level. Following Refs.~\cite{Miyahara:2015cja,Xie:2017erh,Li:2024rqb}, we consider the internal emission mechanism shown in Fig.~\ref{fig:quark}, where the charm quark in the initial $\Lambda_c^+$ weakly decays into a strange quark with the emission of a $W^+$ boson. The $uud$ cluster and a $q\bar{q}$ pair with vacuum quantum numbers hadronize into various meson-baryon pairs. Within the $\Lambda_c^+$ wave function of Refs.~\cite{Capstick:1986ter,Roberts:2007ni}, the weak decay and subsequent hadronization can be expressed as, 
	\begin{align}
		\Lambda_c^+&=\frac{1}{\sqrt{2}}c\left(ud-du\right)\chi_{MA}\nonumber\\ &\Rightarrow\bar{K}^0\frac{1}{\sqrt{2}}u\left(\bar{u}u+\bar{d}d+\bar{s}s\right)\left(ud-du\right)\chi_{MA}, 
		\label{eq:1}
	\end{align}
 where $\chi_{MA}$ denotes the mixed antisymmetric spin wave function of the $(ud-du)$ diquark. Using the pseudoscalar meson matrix $P$ with the $\eta$, $\eta'$ mixing following Ref.~\cite{Bramon:1992kr},
	\begin{eqnarray}
		P =\left(\begin{matrix}  \frac{{\eta}}{\sqrt{3}}+\frac{{\pi}^0}{\sqrt{2}} +\frac{{\eta}'}{\sqrt{6}}  & \pi^+  & K^{+}  \\
			\pi^-  &    \frac{{\eta}}{\sqrt{3}}- \frac{{\pi}^0}{\sqrt{2}}+\frac{{\eta}'}{\sqrt{6}}  &  K^{0} \\
			K^{-}  &  \bar{K}^{0}   &   \sqrt{\frac{2}{3}}{\eta}'-\frac{{\eta}}{\sqrt{3}}
		\end{matrix}
		\right),
	\end{eqnarray}
	the hadronization process in Eq.~\eqref{eq:1} can be further expressed as,
	\begin{align}
		\Lambda_c^+&\Rightarrow\frac{1}{\sqrt{2}}\bar{K}^0\displaystyle \sum_{i}P_{1i}q_{i}\left(ud-du\right)\chi_{MA}\nonumber\\
		&\Rightarrow\frac{1}{\sqrt{2}}\bar{K}^0\Bigg\{\frac{1}{\sqrt{3}}\eta      u(ud-du)+\frac{1}{\sqrt{2}}\pi^0u(ud-du)\nonumber\\
		&+\pi^+d(ud-du)+K^+s(ud-du)\Bigg\} \chi_{MA}\nonumber\\
		&\Rightarrow\bar{K}^0\frac{1}{\sqrt{2}}\Bigg\{\pi^+n+\frac{1}{\sqrt{2}}\pi^0p+\frac{1}{\sqrt{3}}\eta p-\frac{2}{\sqrt{6}}K^+\Lambda \Bigg\}.
		\label{eq:1690MBchannel}
	\end{align}
The baryon octet wave function is given by,
		\begin{align}
	\psi=\frac{1}{\sqrt{2}}(\phi_{MS}\chi_{MS}+\phi_{MA}\chi_{MA}),
		\label{eq:}
	\end{align}
	where $\phi_{MS}\chi_{MS}$ and $\phi_{MA}\chi_{MA}$ denote the mixed-symmetric and mixed-antisymmetry flavor-spin wave functions. The wave function $\phi_{MA}$ is taken from Table III of Ref.~\cite{Miyahara:2016yyh}. Using the isospin triplet $(-\pi^+, \pi^0, \pi^-)$ and the isospin doublet $(p, n)$~\cite{Close:1979bt}, we transform the charge channels in Eq.~\eqref{eq:1690MBchannel} into isospin channels, obtaining,
	\begin{align}
		\Lambda_c^+\Rightarrow\bar{K}^0\frac{1}{\sqrt{2}}\Bigg\{-\frac{\sqrt{6}}{2}\pi N+\frac{\sqrt{3}}{3}\eta N-\frac{\sqrt{6}}{3}K\Lambda \Bigg\},
		\label{eq:5}
	\end{align}
where we have obtained the final states at tree level and incorporated the rescattering effects from channels $\pi N\to\pi N$, $\eta N \to \pi N$, and $K\Lambda\to\pi N$. These rescattering effects are calculated using the chiral unitary approach, with the corresponding hadron-level diagram depicted in Fig.~\ref{fig:hadron1690}.
 
\begin{figure}[tbhp]
		\centering
		\includegraphics[scale=0.8]{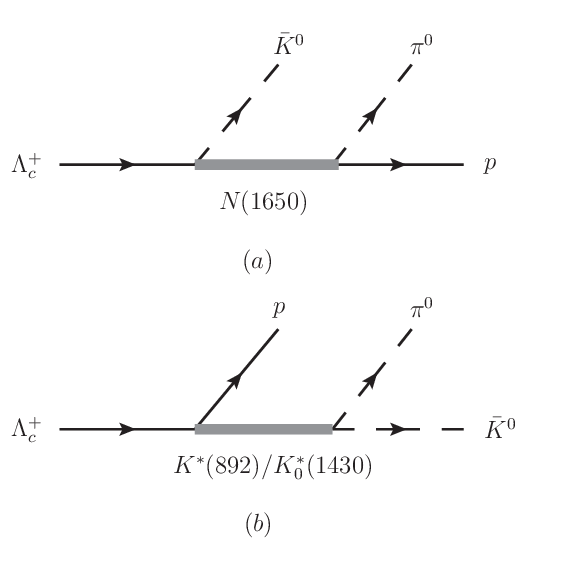}
		\caption{Resonance contributions to $\Lambda_c^+ \to p\bar{K}^0 \pi^0$: (a) $N(1650)$ production, and (b) $K^*(892)/K_0^*(1430)$ production.}
		\label{fig:hadron-N(1650)K^*(892)K_0^*(1430)}
	\end{figure}

	Furthermore, in the results reported by the Belle Collaboration~\cite{Belle:2025voy}, distinct peak structures corresponding to $N(1650)$ and $K^*(892)$ are observed in the $\pi^0p$ and $K_S^0\pi^0$ mass distributions, respectively. Additionally, the threshold enhancement around 1.3~$\mathrm{GeV}$ in the $K_S^0\pi^0$ mass distribution may originate from contribution of the $K_0^*(1430)$. Therefore, we  also consider the contributions from processes $\Lambda_c^+\to N(1650)\bar{K}^0\to p\pi^0\bar{K}^0$ and $\Lambda_c^+\to K^*(892)p/K_0^*(1430)p\to p\pi^0\bar{K}^0$, as illustrated in Fig.~\ref{fig:hadron-N(1650)K^*(892)K_0^*(1430)}.
	
\subsection{Amplitude for Dynamically Generated $N(1535)$}\label{susec:A}

	The dynamical generation mechanism of the $N(1535)$ resonance is shown in Fig.~\ref{fig:hadron1690}, with the corresponding decay amplitude given by,
	\begin{align}
	\mathcal{T}^{N(1535)}=&V_{N(1535)}\bigg[h_{\pi N}+h_{\pi N}G_{\pi N}(M_{\mathrm{inv}})t_{\pi N \to \pi N}(M_{\mathrm{inv}}) \nonumber \\
	&+h_{\eta N}G_{\eta N}(M_{\mathrm{inv}})t_{\eta N \to \pi N}(M_{\mathrm{inv}})
	\nonumber \\
	&+h_{K\Lambda}G_{K\Lambda}(M_{\mathrm{inv}})t_{K\Lambda \to \pi N}(M_{\mathrm{inv}})\bigg],  
	\label{eq:T1535}
\end{align}
where $M_{\mathrm{inv}}$ is the $\pi^0 p$ invariant mass,  $V_{N(1535)}$ parameterizes the production strength of $N(1535)$, and the coefficients $h_{MB}$ ($h_{\pi N}=-\frac{\sqrt{3}}{2}$, $h_{\eta N}=\frac{\sqrt{6}}{6}$, and $h_{K\Lambda}=-\frac{\sqrt{3}}{3}$) from Eq.~\eqref{eq:5} originate from the flavor structure.  

In Eq.~\eqref{eq:T1535}, $G_{MB}$ is the meson-baryon loop function:
\begin{equation} 
G_{MB}(s) = i\int \frac{d^4 q}{(2\pi)^4} \frac{2M_i}{(p - q)^2 - M_i^2 + i\epsilon} \frac{1}{q^2 - m_i^2 + i\epsilon},
\end{equation}
which is evaluated using cut-off regularization with $q_{\mathrm{max}}=1150\mathrm{MeV}$, as taken from Ref.~\cite{Li:2024rqb}. Noting that the threshold of the $\pi N$ channel lies far from the pole position of $N(1535)$, we adopt the following form, as in Ref.~\cite{Li:2025msk,Oset:2024fbk,Dai:2020yfu}.
\begin{equation} 
	G_{\pi N}(s) = 0-i\frac{2M}{8\pi\sqrt{s}}\times\frac{\lambda\left({\rm Re}(s),m^2,M^2\right)}{2\sqrt{s}},
\end{equation}
where $\lambda(x,y,z) = x^2 + y^2 + z^2 - 2xy - 2xz - 2yz$.

 The transition amplitudes $t_{MB\to \pi N}$ for the coupled channels $\pi N$, $\eta N$, $K\Lambda$, and $K\Sigma$ are obtained by solving the Bethe-Salpeter equation,
 \begin{equation} 
 T = [1-VG]^{-1}V,
 \end{equation}
with $V_{ij}$ the interaction kernel obtained from the leading-order chiral Lagrangian, 
	\begin{align}
	V_{ij}=&-C_{ij}\frac{1}{4f_if_j}(2\sqrt{s}-M_i-M_j) \times \nonumber\\
	&\sqrt{\frac{E_i+M_i}{2M_i}}\sqrt{\frac{E_j+M_j}{2M_j}};\\
	f_\pi=93~\mathrm{MeV}&,\qquad f_K=1.2f_\pi,\qquad f_\eta=1.3f_\pi,\nonumber
\end{align} 
   where $M_i/E_i$ and $M_j/E_j$ are the masses and energies of the baryon in the meson-baryon center of mass frame, respectively. The coefficients $C_{ij}$, obtained from the chiral Lagrangians and reflecting $\mathrm{SU(3)}$ flavor symmetry, are given in Table I of Ref.~\cite{Garzon:2014ida}.

\subsection{Contributions from Other Resonances}
	
	We also take into account the contributions from  $S$-wave $N(1650)$, $K_0^*(1430)$, and $P$-wave $K^*(892)$ shown in Fig.~\ref{fig:hadron-N(1650)K^*(892)K_0^*(1430)}, with their corresponding amplitudes,
	\begin{align}
		\mathcal{T}^{N(1650)}=\frac{V_{N(1650)}M_{N(1650)}\Gamma_{N(1650)}}{M_{p\pi^0}^2-M_{N(1650)}^2+iM_{N(1650)}\Gamma_{N(1650)}},  
		\label{eq:T1650}
	\end{align}
    \begin{align}
    \mathcal{T}^{K^*(892)} =& - V_{K^*(892)} \frac{\widetilde{P}_{\pi^0}}{(\widetilde{P}_{\pi^0})_{\text{ave}}} \frac{M_{K^*(892)}}{\pi} \cos\theta_1 \nonumber\\
    &\times \text{Im} \frac{1}{M_{\bar{K}^0\pi^0} - M_{K^*(892)} + i \frac{\Gamma_{K^*(892)}}{2}},  
    \label{eq:T892}
    \end{align}
		\begin{align}
		\mathcal{T}^{K_0^*(1430)}=\frac{V_{K_0^*(1430)}M_{K_0^*(1430)}\Gamma^0_{K_0^*(1430)}}{M_{\bar{K}^0\pi^0}^2-M_{K_0^*(1430)}^2+iM_{K_0^*(1430)}\widetilde{\Gamma}_{K_0^*(1430)}},  
		\label{eq:T1430}
	\end{align}
	where the free parameters $V_X$ represent the production strength of resonance $X$, to be fitted to the Belle data~\cite{Belle:2025voy}. $M_X$ and $\Gamma_X$ are mass and width of the resonance $X$, taken from the central values in Review of Particle Physics (RPP)~\cite{ParticleDataGroup:2024cfk}. In Eq.~\eqref{eq:T892}, $\widetilde{P}_{\pi^0}$ is the $\pi^0$ momentum in the $\bar{K}^0\pi^0$ rest frame, and the average momentum $(\widetilde{P}_{\pi^0})_{\text{ave}}$ is taken as $(\widetilde{P}_{\pi^0})_{\text{ave}}=(M_{\bar{K}^0\pi^0}^{\mathrm{min}}+M_{\bar{K}^0\pi^0}^{\mathrm{max}})/2$. The angle $\theta_1$  between the $\pi^0$ and $p$ momenta in the $\bar{K}^0\pi^0$ rest frame has cosine~\cite{Wang:2015pcn,Wang:2022nac},
	\begin{align}
	\cos\theta_1 = \frac{M_{p\bar{K}^0}^2 - M_{\Lambda_c^+}^2 - m_{\pi^0}^2 + 2\widetilde{P}_{\Lambda_c^+}^0 \widetilde{P}_{\pi^0}^0}{2\widetilde{P}_p \widetilde{P}_{\pi^0}},  
		\label{eq:costheta}
	\end{align}   
	  where $\widetilde{P}_{\Lambda_c^+}^0$ $(\widetilde{P}_{\pi^0}^0)$ is the energy of $\Lambda_c^+$ $(\pi^0)$ in the $\bar{K}^0\pi^0$ rest frame, and $\widetilde{P}_p=\widetilde{P}_{\Lambda_c^+}$ is the proton momentum in this same frame,
\begin{align}
\tilde{P}_{\Lambda_c^+}
&= \frac{\lambda^{1/2}\left(M_{\Lambda_c^+}^2, M_{\bar{K}^0\pi^0}^2, M_{p}^2\right)}{2M_{\bar{K}^0\pi^0}}
= \tilde{P}_{p}, \\[4pt]
\tilde{P}^{0}_{\Lambda_c^+}
&= \sqrt{M_{\Lambda_c^+}^2+\tilde{P}_{\Lambda_c^+}^{\,2}}, \\[4pt]
\tilde{P}_{\pi^0}
&= \frac{\lambda^{1/2}\left(M_{\bar{K}^0\pi^0}^2, m_{\pi^0}^2, m_{\bar{K}^0}^2\right)}{2M_{\bar{K}^0\pi^0}}, \\[4pt]
\tilde{P}^{0}_{\pi^0}
&= \frac{M_{\bar{K}^0\pi^0}^2+m_{\pi^0}^2-m_{\bar{K}^0}^2}{2\,M_{\bar{K}^0\pi^0}}.
\end{align}
	
	For $K_0^*(1430)$, which lies above the $\bar{K}^0\pi^0$ threshold with a width $\Gamma^0_{K_0^*(1430)}=270~\mathrm{MeV}$~\cite{ParticleDataGroup:2024cfk}, we incorporate the Flatté effect~\cite{Lyu:2025aqa},
	\begin{align}
		\widetilde{\Gamma}_{K_0^*(1430)} = \widetilde{\Gamma}_{K\pi}+\widetilde{\Gamma}_{K\eta},  
		\label{eq:widetilde-GammaK_0^*(1430)}
	\end{align}   
	with 
	\begin{align}
		\widetilde{\Gamma}_{K\pi} = \frac{\Gamma_{K\pi}^0}{p_\pi^0} \tilde{p}_\pi \Theta(M_{\bar{K}^0\pi^0} - M_K - M_\pi),  
		\label{eq:widetilde-GammaKpi}
	\end{align}   
		\begin{align}
		\widetilde{\Gamma}_{K\eta} = \frac{\Gamma_{K\eta}^0}{p_\eta^0} \tilde{p}_\eta \Theta(M_{\bar{K}^0\pi^0} - M_K - M_\eta),  
		\label{eq:widetilde-GammaKeta}
	\end{align}   
	where $\Gamma_{K\pi}^0 = 93\%\times\Gamma_{K_0^*(1430)} = 251.1\,\mathrm{MeV}$ and $\Gamma_{K\eta}^0 = 7\%\times\Gamma_{K_0^*(1430)} = 18.9\,\mathrm{MeV}$.

	\subsection{Invariant Mass Distributions}\label{susec:C}

	With the above formalism, one can write down the total amplitude for the process $\Lambda_c^+ \to p\bar{K}^0 \pi^0$ as
	\begin{align}
		\mathcal{T}^{\mathrm{Model A}} =~ &\mathcal{T}^{N(1535)}+\mathcal{T}^{N(1650)}e^{i\phi_{1}}\nonumber\\
		&+\mathcal{T}^{K^*(892)}e^{i\phi_{2}}+\mathcal{T}^{K_0^*(1430)}e^{i\phi_{3}}.
		\label{eq:Total T}
	\end{align}
Using the standard form  for  the three-body decay width from RPP, we have
	\begin{eqnarray}
		\frac{d^2\Gamma}{dM^2_{\pi^0p}{dM^2_{\bar{K}^0p}}}=\frac{4M_{\Lambda_c^+}M_{p}}{{(2\pi)}^3{32M_{\Lambda_c^+}}^3}|\mathcal{T}^{\mathrm{Model A}}|^2, \label {eq:dgamma1} 
	\end{eqnarray}
	\begin{eqnarray}
	\frac{d^2\Gamma}{dM^2_{\bar{K}^0\pi^0}{dM^2_{\pi^0p}}}=\frac{4M_{\Lambda_c^+}M_{p}}{{(2\pi)}^3{32M_{\Lambda_c^+}}^3}|\mathcal{T}^{\mathrm{Model A}}|^2. \label {eq:dgamma2} 
\end{eqnarray}	
The invariant mass distributions $d\Gamma/dM_{\pi^0p}$, $d\Gamma/dM_{p\bar{K}^0}$, and $d\Gamma /dM_{\bar{K}^0\pi^0}$ are obtained by integrating over the other invariant mass variable within the kinematic limits.	
	
\section{Numerical Results and Discussion} \label{sec:Results}
	
\begin{table}[htbp]
    \centering
    \caption{Fitted parameters.}\label{tab:parameters}
    \begin{tabular}{ccc}
        \hline\hline
        Parameter & Model A & Model B \\
        \hline
        $V_{N(1535)}$ & $10.50\pm2.55$ & $6.75\pm6.61$ \\
        $V_{N(1650)}$ & $3.04\pm2.72$ & $7.33\pm3.26$ \\
        $V_{K^*(892)}$ & $5.06\pm2.07$ & $4.75\pm2.06$ \\
        $V_{K_0^*(1430)}$ & $18.72\pm4.19$ & $10.54\pm8.01$ \\
        $V_{N(1440)}$ & --- & $15.71\pm6.08$ \\
        $V_{\Sigma(1750)}$ & --- & $8.68\pm3.46$ \\
        $\phi_{1}$ & $(1.37\pm0.20)\pi$ & $(0.86\pm0.41)\pi$ \\
        $\phi_{2}$ & $(1.97\pm0.29)\pi$ & $(1.51\pm0.41)\pi$ \\
        $\phi_{3}$ & $(0.65\pm0.13)\pi$ & $(-0.25\pm0.51)\pi$ \\
        $\phi_{4}$ & --- & $(-0.43\pm0.40)\pi$ \\
        $\phi_{5}$ & --- & $(1.89\pm0.38)\pi$ \\
        $\chi^2/d.o.f$ & 7.99 & 2.65 \\
        \hline\hline
    \end{tabular}
\end{table}

	\begin{figure}[htbp]
		\centering
		\includegraphics[scale=0.57]{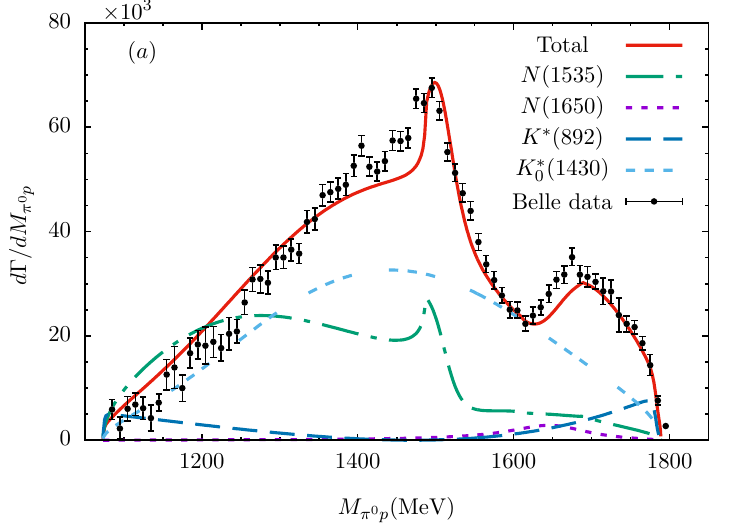}
		\includegraphics[scale=0.57]{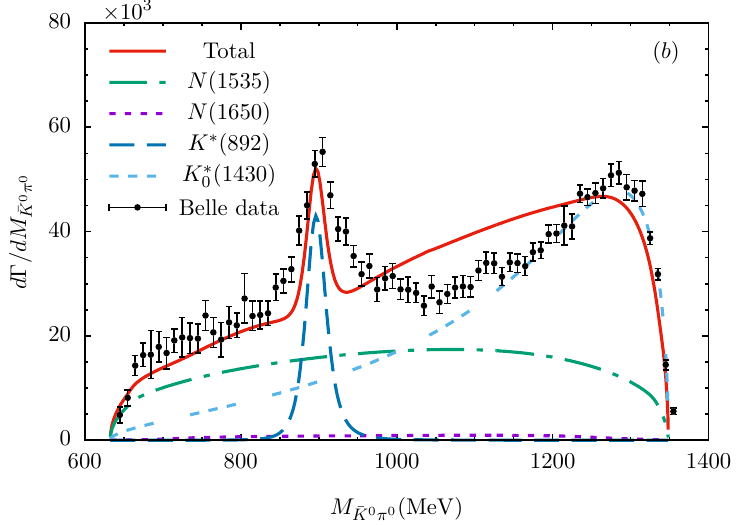}
		\includegraphics[scale=0.57]{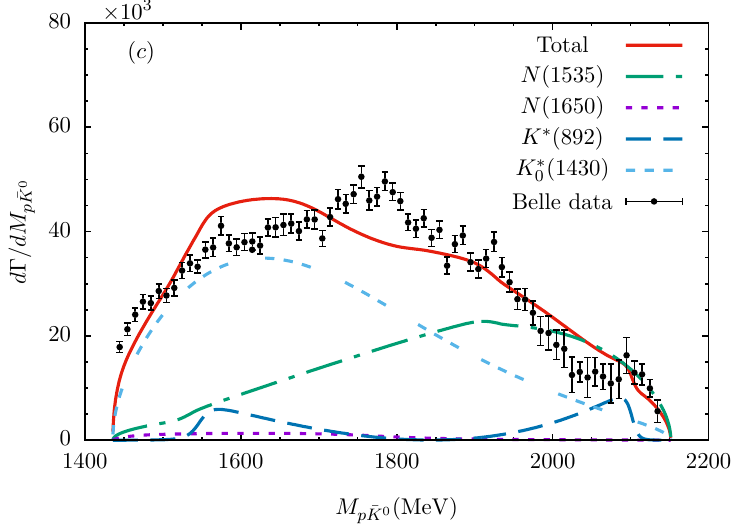}
		\caption{$\pi^0p$ (a), $\bar{K}^0\pi^0$ (b), and $p\bar{K}^0$ (c) invariant mass distributions of the process $\Lambda^+_c\to p\bar{K}^0\pi^0$ with the fitted parameters of Model A. Black data points with error bars labeled by ``Belle data" represent the Belle data taken from Ref.~\cite{Belle:2025voy}, and solid-red curves labeled by ``Total" show the total theoretical results. In addition, we have also presented the contributions from the $N(1650)$, $K^*(892)$, and $K_0^*(1430)$, which are labeled by $N(1650)$, $K^*(892)$, and $K_0^*(1430)$, respectively.}	\label{fig:dgdm-4}
	\end{figure}
	
	\begin{figure}[htbp]
		\centering
		\includegraphics[scale=0.57]{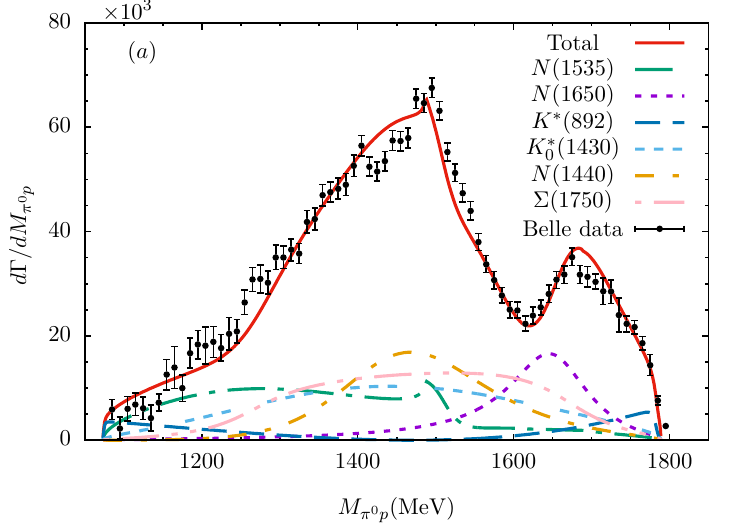}
		\includegraphics[scale=0.57]{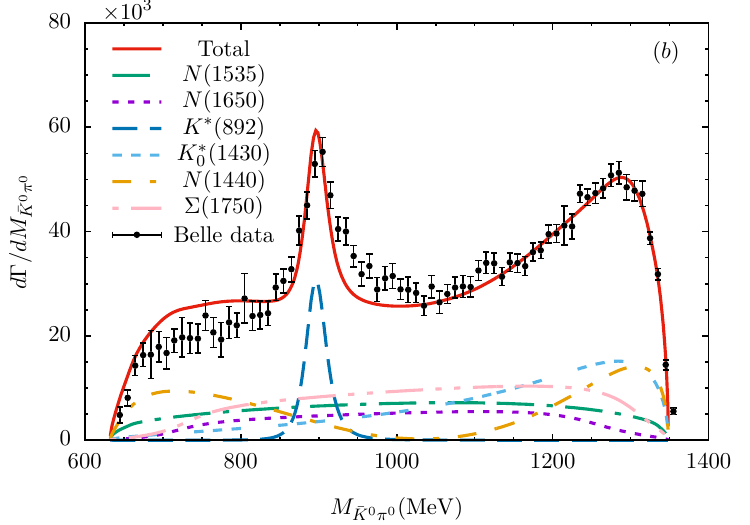}
		\includegraphics[scale=0.57]{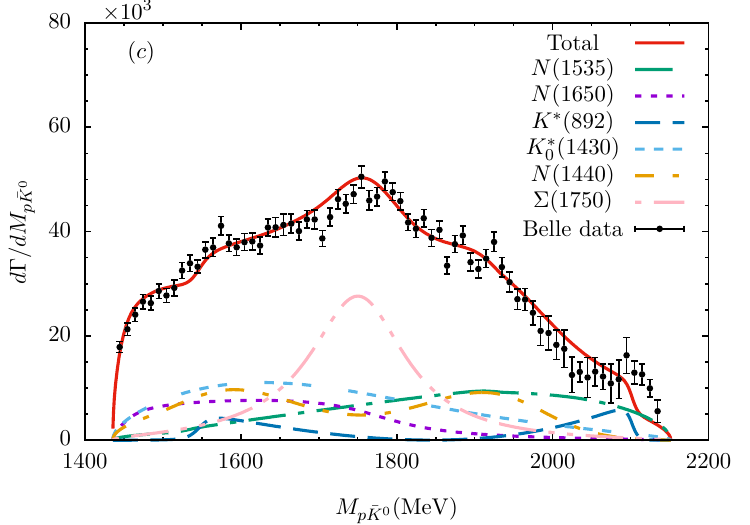}
		\caption{Same as Fig.~\ref{fig:dgdm-4} but with the fitted parameters of Model B including the $N(1440)$ and $\Sigma(1750)$ contributions.}	\label{fig:dgdm-6}
	\end{figure}
	
	\begin{figure}[htbp]
		\centering
		\includegraphics[scale=0.85]{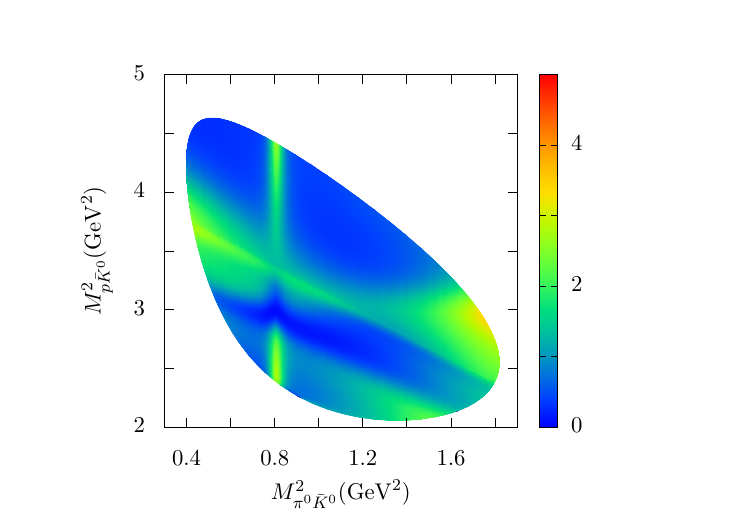}
		\caption{Dalitz plot of the process $\Lambda_c^+ \to p\bar{K}^0 \pi^0$ with the fitted parameters of Model B.}	\label{fig:dalitz}
	\end{figure}

Our model (Model A) contains seven free parameters: $V_{N(1535)}$, $V_{N(1650)}$, $V_{K^*(892)}$, and $V_{K_0^*(1430)}$ representing relative weights, and $\phi_{1}$, $\phi_{2}$, $\phi_{3}$ denoting the interference phases. We fit these parameters to Belle's invariant mass distributions~\cite{Belle:2025voy}, and present the fitted parameters in Table~\ref{tab:parameters} (Model A).

Fig.~\ref{fig:dgdm-4} shows our numerical results for $\pi^0p$, $\bar{K}^0\pi^0$, and $p\bar{K}^0$ mass distributions using the fitted parameters. A clear peak around $1510~\mathrm{MeV}$ in the $\pi^0p$ invariant mass distribution is associated with the dynamically generated $N(1535)$ resonance. Simultaneously, a peak structure appears around $1650~\mathrm{MeV}$, which is due to the $N(1650)$ contribution. Additionally, the $K^*(892)$ contribution reproduces the peak around $900~\mathrm{MeV}$ in the $\bar{K}^0\pi^0$ invariant mass distribution, while $K_0^*(1430)$ accounts for the threshold enhancement structure in the high-energy region. 
	
 While the present model shows overall agreement with experimental data, discrepancies persist around $1750~\mathrm{MeV}$ in the $p\bar{K}^0$ invariant mass distribution, and  $1000-1200~\mathrm{MeV}$ in the $\bar{K}^0\pi^0$ invariant mass distribution. These suggest missing contributions from additional resonances. 
  Although the Belle data show no clear evidence for extra resonances, the $p\bar{K}^0$ region could contain multiple $\Sigma^*$ candidate states in the RPP.

Thus, we extend our model (Model A) to include $P$-wave $N(1440)$ and $S$-wave $\Sigma(1750)$ contributions (Model B), and the corresponding amplitudes can be written as,
 \begin{align}
    \mathcal{T}^{N(1440)} =& - V_{N(1440)} \frac{\widetilde{P}_{\pi^0}}{(\widetilde{P}_{\pi^0})_{\text{ave}}} \frac{M_{N(1440)}}{\pi} \cos\theta_2 \nonumber\\
    &\times \text{Im} \frac{1}{M_{p\pi^0} - M_{N(1440)} + i \frac{\Gamma_{N(1440)}}{2}},  
    \label{eq:T1440}
    \end{align}	 
   \begin{align}	   \mathcal{T}^{\Sigma(1750)}=\frac{V_{\Sigma(1750)}M_{\Sigma(1750)}\Gamma_{\Sigma(1750)}}{M_{p\bar{K}^0}^2-M_{\Sigma(1750)}^2+iM_{\Sigma(1750)}\Gamma_{\Sigma(1750)}}, 
		\label{eq:T1750}
	\end{align}
where $\theta_2$ is the angle between $\pi^0$ and $\bar{K}^0$ in the $\pi^0p$ rest frame. The $\cos\theta_2$ is expressed as
 \begin{align}
	\cos\theta_2 = \frac{M_{\bar{K}^0p}^2 - M_{\Lambda_c^+}^2 - m_{\pi^0}^2 + 2\widetilde{P}_{\Lambda_c^+}^0 \widetilde{P}_{\pi^0}^0}{2\widetilde{P}_{\pi^0} \widetilde{P}_{\bar{K}^0}}.  
		\label{eq:costheta2}
	\end{align}   
\begin{align}
\tilde{P}_{\Lambda_c^+}
&= \frac{\lambda^{1/2}\left(M_{\Lambda_c^+}^2, M_{\pi^0p}^2, m_{\bar{K}^0}^2\right)}{2M_{\pi^0p}}
= \tilde{P}_{\bar{K}^0}, \\
\tilde{P}^{0}_{\Lambda_c^+}
&= \sqrt{M_{\Lambda_c^+}^2+\tilde{P}_{\Lambda_c^+}^{\,2}}, \\
\tilde{P}_{\pi^0}
&= \frac{\lambda^{1/2}\left(M_{\pi^0p}^2, m_{\pi^0}^2, M_{p}^2\right)}{2M_{\pi^0p}}, \\
\tilde{P}^{0}_{\pi^0}
&= \frac{M_{\pi^0p}^2+m_{\pi^0}^2-M_{p}^2}{2\,M_{\pi^0p}}.
\end{align}

 Then, one can write down the total amplitude for Model B as
		\begin{align}
		&\mathcal{T}^{\mathrm{Model B}} =~ \mathcal{T}^{N(1535)}+\mathcal{T}^{N(1650)}e^{i\phi_{1}}+\mathcal{T}^{K^*(892)}e^{i\phi_{2}}\nonumber\\
		&+\mathcal{T}^{K_0^*(1430)}e^{i\phi_{3}}+\mathcal{T}^{N(1440)}e^{i\phi_{4}}+\mathcal{T}^{\Sigma(1750)}e^{i\phi_{5}},
		\label{eq:Total T6}
	\end{align}
    where parameters $\phi_4$ and $\phi_5$ are phase angles. $\mathcal{T}^{\Sigma(1750)}$ and $\mathcal{T}^{N(1440)}$ correspond to the amplitudes of $\Sigma(1750)$ and $N(1440)$, respectively. Now, there are eleven parameters, and the fitted parameters for Model B are summarized in Table~\ref{tab:parameters}. Model B achieves a $\chi^2/\mathrm{d.o.f}=2.65$, which is better than that of Model A $(\chi^2/\mathrm{d.o.f}=7.99)$.

	We have re-plotted the invariant mass distributions with the fitted parameters of Model B in Fig.~\ref{fig:dgdm-6}. The region near $1750~\mathrm{MeV}$ in the $p\bar{K}^0$ invariant mass distribution is well reproduced by $\Sigma(1750)$, and $N(1440)$ significantly improves the description of the $\bar{K}^0\pi^0$ invariant mass distribution.
		
	Then, we present the double differential decay width $d^2\Gamma/(dM_{\pi^0\bar{K}^0}^2dM_{p\bar{K}^0}^2)$ for $\Lambda_c^+ \to p\bar{K}^0 \pi^0$ decay in the $(M_{\pi^0\bar{K}^0}^2, M_{p\bar{K}^0}^2)$ plane with the fitted parameters of Model B in Fig.~\ref{fig:dalitz}, which is in good agreement with the Belle measurements~\cite{Belle:2025voy}.

	\section{Summary} \label{sec:Conclusions}

Recently, the Belle Collaboration has measured the branching fraction ratio of the $\Lambda_c^+\to pK_S^0\pi^0$ decay relative to the $\Lambda_c^+\to pK^-\pi^+$ reaction and reported the final-state invariant mass distributions, in which distinct signals corresponding to the $N(1535)$, $N(1650)$, $K^*(892)$, and $K_0^*(1430)$ resonances can be clearly identified.

We have performed a detailed theoretical analysis of $\Lambda_c^+\to p\bar{K}^0\pi^0$ decay within the chiral unitary approach. Our dynamically generated $N(1535)$ resonance successfully reproduces the peak structure in the $M_{p\pi^0}$ distribution observed by Belle. The inclusion of $N(1650)$, $K^*(892)$, and $K_0^*(1430)$ provides a reasonable description of other features in the invariant mass spectra.

 The extended Model B, incorporating the $N(1440)$ and $\Sigma(1750)$ contributions, significantly improves the agreement with the experimental data, particularly in the $1750$~MeV region of $M_{p\bar{K}^0}$ and the $1000\sim 1200$~MeV region of $M_{\bar{K}^0\pi^0}$. This demonstrates the importance of these additional resonances for a complete description of the decay dynamics.

 Our analysis is consistent with the interpretation of $N(1535)$ as a dynamically generated state from coupled-channel meson-baryon interactions. The successful description of Belle data within this framework provides support for the molecular nature of $N(1535)$.

 Future partial-wave analyses of $\Lambda_c^+\to pK_S^0\pi^0$ by experimental collaborations will further elucidate contributions from intermediate resonances including $K^*$, $\Lambda^*$, $\Sigma^*$, and $\Delta^*$ states. Such analyses, combined with improved theoretical models, will enhance our understanding of non-factorizable processes in charmed baryon decays and provide stringent tests of isospin symmetry.

  \section*{Acknowledgments}
This work is supported by the National Key R\&D Program of China (No. 2024YFE0105200 and 2023YFA1606703), and it is also supported by the Natural Science Foundation of Henan under Grant No. 232300421140, and by the National Natural Science Foundation of China under Grants Nos. 12475086, 12192263, 12575094, W2543006, 12435007 and 12361141819.

	
\end{document}